\documentclass[twocolumn,prl,showpacs,superscriptaddress]{revtex4-1}
\usepackage[latin9]{inputenc}
\usepackage{amsmath}
\usepackage{amssymb}
\usepackage{esint}

\makeatletter
\@ifundefined{textcolor}{}
{%
 \definecolor{BLACK}{gray}{0}
 \definecolor{WHITE}{gray}{1}
 \definecolor{RED}{rgb}{1,0,0}
 \definecolor{GREEN}{rgb}{0,1,0}
 \definecolor{BLUE}{rgb}{0,0,1}
 \definecolor{CYAN}{cmyk}{1,0,0,0}
 \definecolor{MAGENTA}{cmyk}{0,1,0,0}
 \definecolor{YELLOW}{cmyk}{0,0,1,0}
}

\usepackage{subfigure}
\usepackage{epsfig}
\usepackage{amsfonts}
\usepackage{mathrsfs}
\usepackage{CJK}
\usepackage[toc,page,title,titletoc,header]{appendix}
\usepackage{physics}

\makeatother

\begin{document}

\title{Configuration dependent reflection induced by dissipated localized modes}

\author{Wei Zhu}

\affiliation{Institute of Physics, Beijing National Laboratory for
  Condensed Matter Physics, Chinese Academy of Sciences, Beijing
  100190, China}

\author{Peng Zhang}

\affiliation{Department of Physics, Renmin University of China,
  Beijing, 100190, China}

\affiliation{Beijing Key Laboratory of Opto-electronic Functional
  Materials \& Micro-nano Devices (Renmin University of China)}

\author{D. L. Zhou}

\email{zhoudl72@iphy.ac.cn}

\affiliation{Institute of Physics, Beijing National Laboratory for
  Condensed Matter Physics, Chinese Academy of Sciences, Beijing
  100190, China}

\begin{abstract}
  We study the one photon scattering problem for a super cavity (SC)
  coupling with two two-level atoms. With atomic decay, we find a
  sudden drop in reflection at $\Delta=0$ for the two atoms in the
  node-antinode configuration but not in the antinode-node one. The
  underlying mechanism is due to the scatterer has a configuration
  dependent localized dissipated eigen-mode at $\Delta=0$. In the
  node-antinode configuration, the eigen-mode localized near the input
  side, which can transport the photon into the SC\@. By exciting the
  atom at the node, the photon can leak into the reservoir due to
  atomic decay, which causes the sudden drop at $\Delta=0$ in
  reflection. In the antinode-node case, however, the eigen-mode is
  localized near the output side, no photon can be transported into
  the SC and leads to completely reflection at $\Delta=0$. A similar
  phenomenon has been observed in a recent experiment of X-ray quantum
  optics [Nature \textbf{482}, 199 (2012)] but with a much more
  complicated explanation due to electromagnetically induced
  transparency.
\end{abstract}

\maketitle

\section{Introduction}
\label{sec:introduction}

The studies on photon scattering play an important role in quantum
optics. Various significant phenomena in quantum optics are related to
scattering, such as electromagnetically induced transparency (EIT) and
resonance fluorescence~\cite{r1,master}. Photon scattering for models
like single or multi atoms coupling with optical
cavity~\cite{r6,xray1,xray2,xray2t} or
waveguide~\cite{r2,r3,r7,1,r8,Fratini14,Dai15} have long long been
active research area both in theory and experiment. In recent years, a
theoretical model based on coupling cavity array (CCA) is proposed and
quickly attracts a great deal of attention~\cite{r9}. CCA is a perfect
platform for scattering study, important achievements have been made
both on the one dimensional~\cite{r9,r10,r11,r12} and the two
dimensional~\cite{peng1} CCA platform.

In nature, the spontaneous emission of the atoms are inevitable due
to their coupling with the surrounding electromagnetic
environment~\cite{r4,r5}, and the atomic decay has long been included
in the scattering research through master equation~\cite{1,2}. But
except the intuitive results as the decrease and expansion of the
reflection (transmission) peaks, no qualitative difference has been
found in the past researches for atoms coupling with waveguide or
CCA\@.

In this paper, we revisit the photon scattering problem for two
two-level atoms coupling with the super cavity~\cite{3} but include
atomic decay. A super cavity (SC) is first present in our former work
to help study the photon scattering for a two-level atom coupling with
a multi-mode cavity on the CCA platform~\cite{zhou1}. In the two atoms
case, whether including the atomic decay or not leads to a fundamental
difference in the scattering results. Without atomic decay, the
scattering results for the two atoms are very similar with the one
atom case, configuration of the atoms has no relevant effect~\cite{3}. When
including the atomic decay, for reflection a dip appears around the
resonant energy ($\Delta=0$) only for the atoms arranged in
node-antinode configuration corresponding to the resonant mode. A very
similar phenomenon has been observed in Ref.~\cite{xray2} and a
complicated theory which attributes it to EIT is given. In this paper,
we reveals the physical mechanism behind this phenomenon is much
simpler and has nothing to do with EIT\@.

The rest of the paper is organized as follows. In Sec.~\ref{sec:2}, we
introduce our model and briefly give the results for the condition
without atomic decay. In Sec.~\ref{sec:3}, we use the master equation
to handle the photon scattering problem with atomic decay. We reveal
the physical mechanism behind the configuration dependent phenomenon. In
Sec.~\ref{sec:4}, single-mode approximation is introduced to support
the analysis we give above. Finally a brief conclusion is given.

\section{Model: two two-level atoms in a super-cavity \label{sec:2}}

\begin{figure}[htbp]
  \includegraphics[width=8cm]{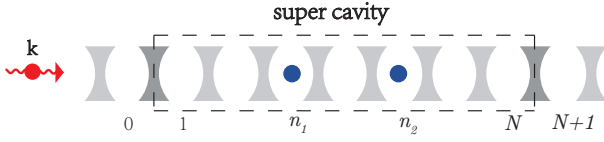}
  \caption{(Color online). Schematic set of the single-photon
    scattering problem for the 1D CCA model. One photon (filled red
    circle) with the wave vector $k$ injects from the left side of the
    SC composed of $N$ cavities. The SC is formed by a
    relatively small coupling strength $\eta$ with the outside
    cavities. Two two-level atoms (filled blue circle) are in the
    $n_{1}$-th and $n_{2}$-th cavities of the SC respectively.\@ Here
    we take $N=7$, $n_{1}=3$ and $n_{2}=5$.\label{fig:1}}
\end{figure}

The system we consider consists of three parts, see Fig.~\ref{fig:1}.
The central part contains a SC with two two-level atoms
embedded in, where the SC is formed by a 1D single-mode
cavity array with $N$ cavities, and these two atoms interacts with two
cavities of the SC respectively. The second (third) part is
the left (right) photon channel, formed by a semi-infinite 1D cavity
array connected to our central part from its left (right) side.

We will study single-photon scattering problem on this system. One
photon with wave vector $k$ from the left photon channel is scattered
by the SC system. Our aim is to figure out how the
reflection and transmission coefficients depend on the position of the
two atoms in the SC. More precisely, we will study the cases
when one atom is at the node of the resonant mode of the SC
while the other is at the antinode. In particular, we are interested
in whether the position order of the two atoms, i.e., the
node-antinode configuration or the antinode-node configuration for the
resonant mode, is related.

Under the rotating wave approximation, the Hamiltonian of our system
is given by
\begin{equation}
  H = H_S + H_L + H_R + H_{SL} + H_{SR},
\end{equation}
where
\begin{align}
  H_{S} & =  \sum^{N}_{j=1} \omega_{c}a^{\dag}_{j}a_{j}
          - \sum^{N}_{j=2} \xi (a^{\dag}_{j-1} a_{j} +
          a^{\dag}_{j} a_{j-1}) \nonumber\\
        & \quad {} + \sum_{i=1}^2 \left[\omega_{a} |e\rangle_{i}
          \langle e| +  \Omega (a^{\dag}_{n_i} \sigma^{-}_{i} +
          \mathrm{H.c.})\right], \label{h0} \\
  H_{L} & =  \sum^{0}_{j=-\infty} \left[ \omega_{c}a^{\dag}_{j}a_{j}
          - \xi (a^{\dag}_{j-1}a_{j} + a^{\dag}_{j}a_{j-1}) \right], \\
  H_{R} & =  \sum^{\infty}_{j=N+1} \left[ \omega_{c}a^{\dag}_{j}a_{j}
          - \xi(a^{\dag}_{j}a_{j+1}+a^{\dag}_{j}a_{j+1}) \right], \\
  H_{SL} & =  -\eta(a^{\dag}_{0}a_{1} + \mathrm{H.c.}),\\
  H_{SR} & =  -\eta(a^{\dag}_{N}a_{N+1}+ \mathrm{H.c.}).\label{hi}
\end{align}
Here $H_{S}$ is the free Hamiltonian for the central part, and $H_{L}$
($H_{R}$) is the Hamiltonian for the left (right) photon channel,
$a_{j}$ ($a_{j}^{\dagger}$) is the annihilation (creation) operator of
the photon in the $j$-th cavity, $|e_{i}\rangle$ ($i=1,2$) is the
excited state of the atom $i$, $n_{i}$ is the label of the cavity that
the atom $i$ interacts with, $\Omega$ the coupling strength between
each atom and its cavity, $\omega_c$ is the mode frequency for each
cavity,  $\omega_a$ is the eigen-energy of the atomic excited state,
$\xi$ is the hopping strength between nearest neighbor cavities within
the three parts, and $\eta$ is the hopping strength between nearest
neighbor cavities between the SC system and the left or right photon
channel.

Without atomic decay, due to the excitation number is conserved in
this model, the scattering state can be expanded as
\begin{equation}
  |\psi_{k}^{(+)}\rangle = \sum_{l} C_{l} |1_{l};g_{1},g_{2}\rangle + \alpha_{1}
  |\text{vac};e_{1},g_{2}\rangle + \alpha_{2}|\text{vac};g_{1},e_{2}\rangle,\label{psik}
\end{equation}
with
\begin{equation}
  C_{l} = \left\{
    \begin{array}{cc}
      e^{ikl}+re^{-ikl}, & l<0,\\
      c_{1}e^{ikl}+d_{1}e^{-ikl}, & 0< l< n_{1},\\
      c_{2}e^{ikl}+d_{2}e^{-ikl}, & n_{1}< l< n_{2},\\
      c_{3}e^{ikl}+d_{3}e^{-ikl}, & n_{2}<l< N, \\
      te^{ikl}, & l< N.
    \end{array}
  \right.\label{cl}
\end{equation}
Here the parameters $t$ and $r$ are the single-photon transmission and
reflection amplitudes, respectively. Meanwhile the wave function
must be continuous at nodes $0$, $n_1$, $n_2$ and $N$. According to
the scattering theory \cite{taylor}, the scattering state
$|\psi_{k}^{(+)}\rangle$ is an eigen-state of the Hamiltonian $H$ with
eigen-energy $E_{k}$, i.e., we have
\begin{equation}
  H|\psi_{k}^{(+)}\rangle=E_{k}|\psi_{k}^{(+)}\rangle.\label{eigen}
\end{equation}

\begin{figure}[htbp]
  \includegraphics[width=1\columnwidth]{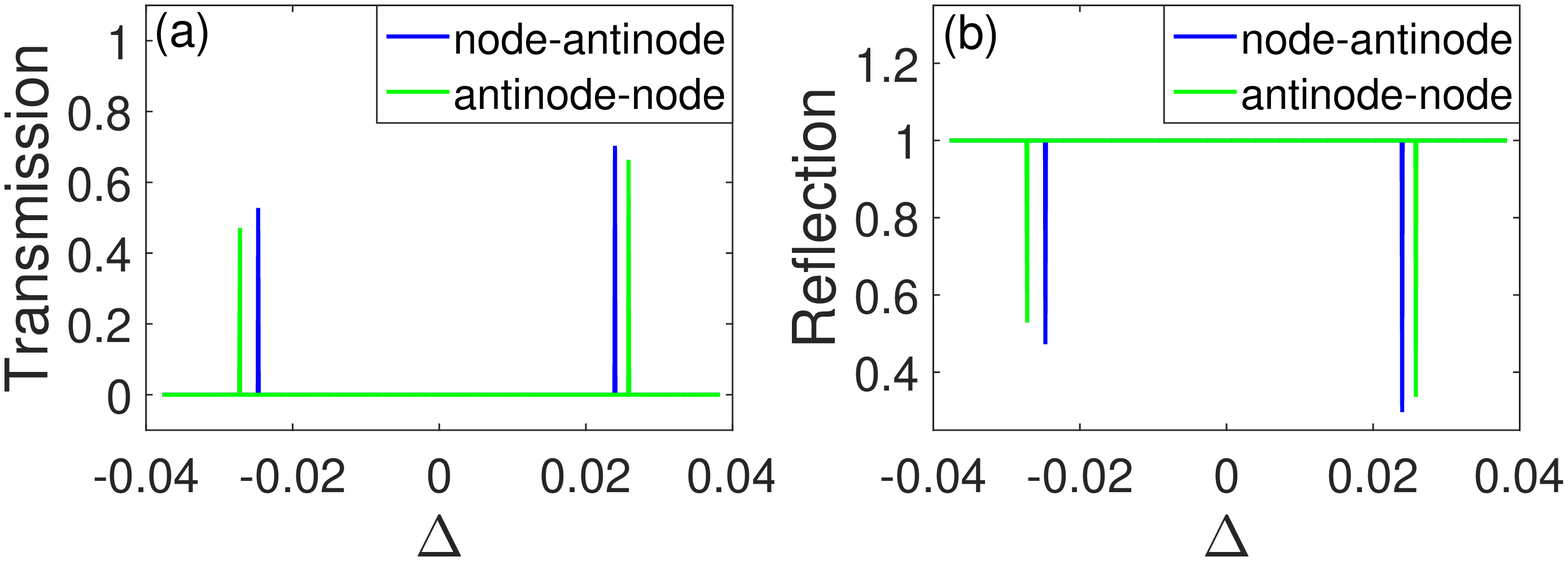}
  \caption{(Color online).(a) Single-photon reflection for system
    without decay. (b) Single-photon transmission for system without
    decay. Blue solid line represents the atoms in node-antinode
    (8-12) configuration while the green solid line is for the
    antinode-node (12-16) configuration. Here, $N=31$, $\eta=0.01$ and
    $\Omega=0.1$.}
  \label{fig:5}
\end{figure}

Numerically solving Eq.~(\ref{eigen}) we can obtain the transmission
($T=|t|^{2}$) and reflection ($R=|r|^{2}$) coefficients. The case
without decay has been thoroughly investigated in our precious work~\cite{3}.
As Fig.~\ref{fig:5} shows, no qualitative difference appears in
transmission or reflection between arranging the atoms in
node-antinode and antinode-node configurations. Actually the result is
quite similar with the one-atom situation~\cite{zhou1}.

\section{Scattering with decay \label{sec:3}}
Now we consider the situation with the atomic decay. Since the two
atoms locate in two distant cavities, it is reasonable to assume
that each atom is coupling with an independent reservoir. Then the
dynamics of our system is controlled by the master
equation~\cite{master}
\begin{align}
  \label{eq:4}
  \dv{\rho(t)}{t} & = -i \comm{H}{\rho(t)} + \frac{\gamma}{2} \sum_{l=1}^{2}
                    \nonumber\\
                  & \quad  \pqty{2 \sigma_{l}^{-} \rho(t)
                    \sigma_{l}^{+} - \op{e_{l}}
                    \rho(t) - \rho(t) \op{e_{l}}},
\end{align}
where $\gamma$ is the spontaneous decay rate of the atomic excited
state. Here we assume the decay rate for each atom is the same.

The steady state for our scattering problem is
\begin{equation}
  \label{eq:5}
  \rho = \op{\Psi} + \kappa \op{G},
\end{equation}
with $\ket{G}=\ket{\mathrm{vac};g_{1},g_{2}}$ being the ground state of
our system, and $\ket{\Psi}$ being the scattering state
\begin{equation}
  \label{eq:2}
  \ket{\Psi} = \ket{k}_{L} + r \ket{-k}_{L} + \sum_{j=1}^{N} c_{j}
  \ket{j} + d_{1} \ket{e_{1}} + d_{2} \ket{e_{2}} + t \ket{k}_{R},
\end{equation}
where
\begin{align}
  \label{eq:3}
  \ket{k}_{L} & = \sum_{j=-\infty}^{0} e^{i k j} \ket{j}, \\
  \ket{k}_{R} & = \sum_{j=N+1}^{\infty} e^{i k j} \ket{j},
\end{align}
with $\ket{j}=a_{j}^{\dagger}\ket{G}$ and
$\ket{e_{l}}=\sigma_{l}^{+}\ket{G}$. Then the time independent master
equation implies that the scattering state $\ket{\Psi}$ satisfies
\begin{equation}
  \label{eq:1}
  \pqty{H - i\frac{\gamma}{2} \sum_{l=1}^{2} \op{e_{l}}} \ket{\Psi} =
  E_{k} \ket{\Psi},
\end{equation}
where $E_{k}=\omega_{c}-2\eta \cos k$. So we can use the effective
Hamiltonian $H_{eff}=H - i\frac{\gamma}{2} \sum_{l=1}^{2} \op{e_{l}}$
to describe this decay system~\cite{1,2}. Note that Eq.~\eqref{eq:1}
is sufficient to determine the transmission coefficient
$T=\abs{t}^{2}$ and the reflection coefficient $R=\abs{r}^{2}$.

\begin{figure}[htbp]
  \includegraphics[width=1\columnwidth]{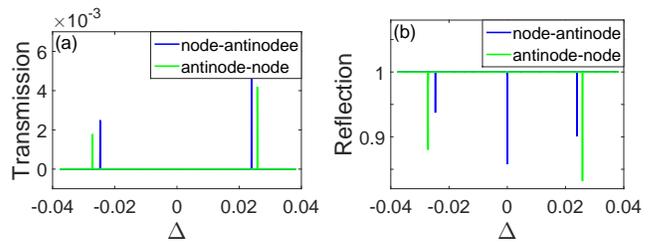}
  \caption{(Color online). Transmission and reflection spectrum for
    the system with decay. (a) Single-photon transmission for two
    atoms in different configurations with decay. Blue solid line
    represents the node-antinode (8-12) configuration while the green
    solid line is for the antinode-node (12-16) configuration. (b)
    Single-photon reflection. Here, $\gamma=10^{-5}$ and the other
    parameters are the same as in Fig.~2.}
  \label{fig:3}
\end{figure}

In Fig.~\ref{fig:3} we give the numerical results for single-photon
transmission and reflection coefficients varying with $\Delta$. Here,
$\Delta=E_{k}-E_{n}$ is the energy difference between the incoming photon
and the resonant mode of the SC\@. For transmission, we see two peaks
in our selected region of $\Delta$ and their space between is
configuration-dependent. This result is quite similar with the case
without atomic decay (see Fig.~\ref{fig:5}), and due to the atomic
decay the transmission has a dramatic decline. As for reflection,
despite the peaks we expect to appear at the same positions
corresponding to the transmission, one more peak emerges or in another
word the reflection experience a sudden drop at $\Delta=0$ only for
the node-antinode configuration. This is a major difference comparing
with the situation without the atomic decay. A qualitatively similar
phenomenon has been observed in Ref.~\cite{xray2}, which is explained
with a much complicated model. Moreover, the influence of the atomic
decay on reflection is much smaller. Next we will try to figure out
the physical mechanism underlying this interesting configuration-dependent
phenomenon.

Similarly as that discussed in Ref.~\cite{zhou1}, the reflection
coefficient $R$ at $E_{k}$ for our setting is essentially determined
by the eigen-modes of $H_{S}$ near resonant with $E_{k}$. The obvious
qualitative difference in $R$ between the two cases for the two atoms
locating in the node-antinode configuration or in the antinode-node
configuration occurs at $\Delta=0$, i.e., the input photon with energy
resonant with the resonant eigen-mode of the empty SC\@. In our model
the scatterer consists of SC and two atoms resonant coupling with SC's
$n$-th eigen-mode (with eigenvalue $E_{n}$), we can prove that $E_{n}$
is also an eigen-value of the scatterer ($H_s$) only if either atom is
located at the node of the mode. Thus, the condition is met for the
above two configurations, and $E_{n}$ will still be an eigen-value of
the scatterer in both situations. But no peak appears at $\Delta=0$
for the reflection or transmission without the atomic decay (see
Fig.~\ref{fig:5}) which violates the resonant tunneling assumption.
Therefore we need to analyze the eigen-mode of the SC system when
$\Delta=0$.

\begin{figure}[htbp]
  \includegraphics[width=1\columnwidth]{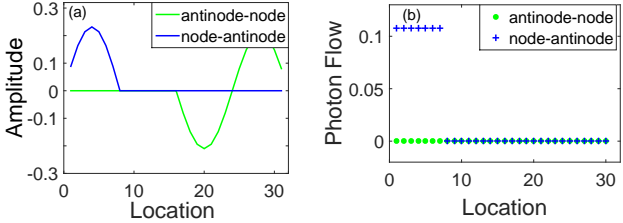}
  \caption{(Color online). (a) Eigen-mode of the scatterer ($H_s$)
    with $E=-2\cos{\frac{4\pi}{N+1}}$ for the atoms in different
    configurations. Blue and green solid lines each stands for
    node-antinode and node-antinode configuration. (b) Photon flow for
    each cavity in the SC at $\Delta=0$. Blue cross (green dot) is for
    the node-antinode (antinode-node) configuration. Here, the
    parameters are the same as in Fig.~\ref{fig:3}. }
  \label{fig:4}
\end{figure}

In order to clarify the confusion, Fig.~\ref{fig:4}(a) shows this
special mode with atoms arranged in two configurations. This mode is
localized and its localization condition is configuration dependent.
As for the node-antinode configuration the mode localized between the
left wall of the SC and the atom at node while for the antinode-node
configuration it localized between the atom at node and the right
wall. The appearance of this localized mode is due to coupling between
the node atom and the non-resonance modes~\cite{zhou1}, and the
antinode atom should not be excited and there are no photons around
it.

Due to the mode is localized, no photon can be transported through the
SC under this incoming energy. So the reflection (transmission) shows
no peak at $\Delta=0$ without the atomic decay, no matter atoms arranged in which
configuration. When the atoms coupling with the reservoir, new photon
leakage way is introduced due to spontaneous emission. Thus mode
localized in different way can lead to fundamental difference. For
antinode-node configuration, the mode is localized at the output
(right) side of the SC so the incoming photon from left to the SC is
totally reflected by the left wall. Then with or without the atomic decay makes
no difference since no photon goes into SC at all, so $R=1$ ($T=0$) at
$\Delta=0$. For the node-antinode configuration, the photon has
probability to go into the SC as the mode localized at the input
(left) side of the SC\@. The incoming photon will excite the atom at node
and then leak into the reservoir due to spontaneous emission. This
leads to the appearance of the sudden drop of reflection at
$\Delta=0$.

Next we calculate the photon flow inside the SC at $\Delta=0$. The
photon flow for the $l$-th cavity is defined as
\begin{equation}
  J_l=-i[C_l(C^{*}_{l+1}-C^{*}_{l})+C_l^{*}(C_{l+1}-C_{l})].\label{flow}
\end{equation}
Fig.~\ref{fig:4}(b) shows no photon flow within the SC for the atoms
arranged in antinode-node configuration, meanwhile for the
node-antinode configuration there is a steady photon flow name $J_s$
before the node atom. This confirms the analysis we give above.
Further, the steady incoming photon is $J_i=2\sin{k}$, the leaking
rate $L$ of the photon into the reservoir is defined as $L=J_s/J_i$.
Then we numerically obtain $L+R+T=1$ as expected. Another major
consequence by introducing the decay is the decline of the
transmission. Under the above mentioned parameter condition
$\gamma=10^{-5}$, the transmission is so small that can be ignored
(see Fig.~\ref{fig:3}). Thus closing the transmission channel by
setting $\eta=0$ for the right wall of the SC will cause little change
to the above results but making our model much similar with the
experiment condition~\cite{xray2}.

The localized eigen-mode can be analytically solved. When the wave
vector of the injection photon is $k=\frac{l\pi}{N+1}$, the atom
$n_{i}$ in the node implies that $\sin(k n_{i})=0$, and the atom in
the antinode implies that $\vert\sin(k n_{i})\vert=1$. For example,
$\sin(k n_{1})=0$ and $\vert\sin(k n_{2})\vert=1$ in the case of
node-antinode configuration. In this case, we find an analytical
solution of the eigen-mode of $H_{s}$ with eigen-value
$E_{l}=\omega_{c}-2\cos k$:
\begin{equation}
  |\psi_{l}\rangle = \sum_{j=1}^{N} b_{j} |1_{j}; g_{1}, g_{2}\rangle
  + \alpha|\mathrm{\text{vac}}; e_{1}, g_{2}\rangle,
\end{equation}
where
\begin{align}
  b_j & =
        \begin{cases}
          \sin(kj) \sqrt{\frac{n_{1}}{2}+\frac{\sin^2{k}}{g^2}},  &
          1\leq j\leq n_{1}, \\
          0, & n_{1}+1\leq j\leq N,
        \end{cases} \\
  \alpha & = - \frac{b_{n_{1}-1}}{g}.
\end{align}
The localized mode for antinode-node configuration is almost the same
only with the state localized to the right.

\section{Single Mode Approximation \label{sec:4}}

For $\Delta=0$, we believe only the localized mode is important and
this is the starting point of our analysis above. To prove the validness of this
assumption, we introduce the single mode approximation Hamiltonian of the scatter
\begin{equation}
  H_{S}=E_l|\psi_{l}\rangle\langle\psi_{l}|
\end{equation}
while keeping $H_{L}$,$H_{R}$,$H_{SL}$ and $H_{SR}$ the same. Now the
scattering state can be expanded as
\begin{equation}
  |\Psi_{k}^{(+)}\rangle = |\varphi_{k}\rangle + r |\varphi_{k}\rangle
  + \mu |\psi_{l}\rangle + t |\phi_k\rangle
\end{equation}
with
\begin{equation}
  \begin{cases}
    |\varphi_{k}\rangle=\sum_{j=-\infty}^{0}e^{ikj}|j\rangle, \\
    |\phi_k\rangle=\sum_{j=N+1}^{\infty}e^{ikj}|j\rangle.
  \end{cases}
\end{equation}
Without the atomic decay, through Eq.~(\ref{eigen}) we have
\begin{eqnarray}
  \Delta^{'}\mu+\eta b_1(1+r)+t\eta b_Ne^{ik(N+1)}=0, \\
  \eta b_1\mu-(e^{ik}+re^{-ik})=0, \\
  \eta b_N\mu-te^{ikN}=0
\end{eqnarray}
with $\Delta^{'}=E_k-E_l$. For the antinode-node configuration, $b_1=0$
leads to $R=|r|^2=1$. While for the node-antinode case, $b_N=0$ leads
to $t=0$ and we can analytically solve $r$ as
\begin{equation}
  r=-\frac{\Delta^{'}e^{ik}+b_1\eta^2}{b_1\eta^2+\Delta^{'}e^{-ik}}
\end{equation}
leading to $R=|r|^2=1$.

\begin{figure}[htbp]
  \includegraphics[width=8cm]{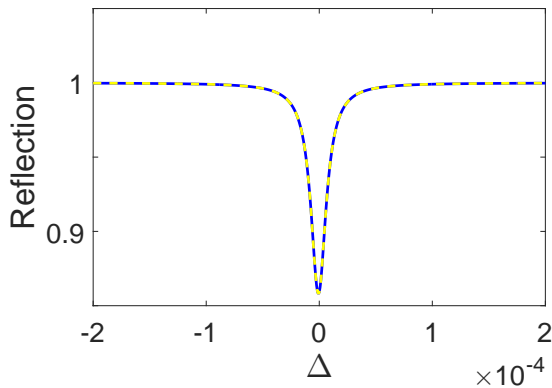}
  \caption{(Color online). Reflection vs $\Delta$ around $\Delta=0$.
    The blue solid line represents the exact numerical result, the
    yellow dashed line is obtained through Eq.~(\ref{single-mode}).
    Here, the parameters are the same as in Fig.~\ref{fig:3}.}
  \label{fig:6}
\end{figure}

For the condition with atomic decay, we use the standard master equation to
handle. The master equation for steady state is
\begin{equation}
  -i[H,\rho] - \frac{\Gamma}{2} (\rho \sigma_{+}^{1} \sigma_{-}^{1} +
  \sigma_{+}^{1} \sigma_{-}^{1} \rho) + \Gamma \sigma_{-}^{1} \rho
  \sigma_{+}^{1} = 0
\end{equation}
with
\begin{equation}
  \rho = \kappa |\psi_{l}\rangle \langle\psi_{l}| + (1 - \kappa)
  |\widetilde{\text{vac}}\rangle \langle\widetilde{\text{vac}}|.
\end{equation}
Projecting the master equation to various bases, we obtain the following
independent equations
\begin{align}
  & (\Delta^{'}+i\Gamma|\alpha|^2/2)\mu \nonumber\\
    & \quad {}+\eta b_1(1+r)+t\eta b_N e^{ik(N+1)}=0, \\
  & \eta b_1\mu-(e^{ik}+re^{-ik})=0, \\
  & \eta b_N\mu-te^{ikN}=0,
\end{align}
which can be analytically solved:
\begin{equation}
  r=\frac{e^{ik}-\eta b_1\beta}{\eta b_1\beta-e^{-ik}}, \label{single-mode}
\end{equation}
where $\beta=\frac{i\eta b_1}{\Gamma|\alpha|^2/2-i\Delta^{'}}$. For
antinode-node configuration, $b_1=0$ and we obtain the same result
$R=1$ as the condition without the atomic decay. Meanwhile for the node-antinode configuration,
Eq.~(\ref{single-mode}) can perfectly give the reflection coefficient
around $\Delta=0$ as show in Fig.~\ref{fig:6}. Thus compare
with the exact model, we obtain the same results near $\Delta=0$ based on this
single-mode approximation model no matter under the condition without
or with atomic decay.

\section{Conclusion}

In conclusion, based on the 1D CCA platform we investigate the
single-photon scattering problem with a SC coupling with two atoms
under the condition with decay. Compared with the condition without
atomic decay, the reflection with decay shows a significant
difference, it will drop suddenly (peak) at $\Delta=0$ only for the
node-antinode configuration. We propose the EIT like phenomenon is
actually not determined by EIT mechanism. It is due to the special
eigen-mode condition for the scatterer at $\Delta=0$. This eigen-mode
is localized and its localization condition is configuration
dependent, so photon can not be transported through this mode. This
explains why without the atomic decay, $R=1$ ($T=0$) at $\Delta=0$ in
any of the two configurations. With atomic decay, the eigen-mode for
node-antinode configuration can transport photon into the SC, and the
photon can leak into the reservoir through exciting the atom at the
node. So the reflection shows a sudden drop at $\Delta=0$. Meanwhile,
no photon can be transported into the SC through the eigen-mode for
antinode-node configuration which leads to $R=1$ at $\Delta=0$. We
calculate the photon flow and use the results of the single-mode
approximation to support our analysis. We hope our analysis can help
understand the experiment results and enlighten the study for using 1D
CCA to simulate real experiment settings.

\begin{acknowledgements}
  The authors thank Y.~Li and C.P.~Sun for helpful discussions. This
  work has been supported by National Natural Science Foundation of
  China under Grants Nos. 11475254, 11222430, 11434011, and NKBRSF of
  China under Grants No. 2014CB921202.
\end{acknowledgements}

\end{document}